\documentclass[journal=ancac3,manuscript=article,layout=twocolumn]{achemso}

\usepackage{chemformula} % Formula subscripts using \ch{}
\usepackage[T1]{fontenc} % Use modern font encodings
\usepackage[utf8]{inputenc}
\usepackage{float}
\usepackage{amsmath,amssymb}

\author{Samantha Sbarra}
\affiliation{Matériaux et Phénomènes Quantiques, Université Paris Cité, CNRS, UMR 7162, 75013 Paris, France}

\author{Louis Waquier}
\affiliation{Matériaux et Phénomènes Quantiques, Université Paris Cité, CNRS, UMR 7162, 75013 Paris, France}

\author{Stephan Suffit}
\affiliation{Matériaux et Phénomènes Quantiques, Université Paris Cité, CNRS, UMR 7162, 75013 Paris, France}

\author{Aristide Lemaître}
\affiliation{Centre de Nanosciences et de Nanotechnologies, Université Paris-Saclay, CNRS, UMR 9001, 91120 Palaiseau, France}

\author{Ivan Favero}
\affiliation{Matériaux et Phénomènes Quantiques, Université Paris Cité, CNRS, UMR 7162, 75013 Paris, France}
\email{ivan.favero@u-paris.fr}
\title[An \textsf{achemso} demo]
{Optomechanical measurement of single nanodroplet evaporation with millisecond time-resolution}

\begin{document}

\begin{abstract}
  
Tracking the evolution of an individual nanodroplet of liquid in real-time remains an outstanding challenge. Here a miniature optomechanical resonator detects a single nanodroplet landing on a surface and measures its subsequent evaporation down to a volume of twenty attoliters. The ultra-high mechanical frequency and sensitivity of the device enable a time resolution below the millisecond, sufficient to resolve the fast evaporation dynamics under ambient conditions. Using the device dual optical and mechanical capability, we determine the evaporation in the first ten milliseconds to occur at constant contact radius with a dynamics ruled by the mere Kelvin effect, producing evaporation despite a saturated surrounding gas. Over the following hundred of milliseconds, the droplet further shrinks while being accompanied by the spreading of an underlying puddle. In the final steady-state after evaporation, an extended thin liquid film is stabilized on the surface. Our optomechanical technique opens the unique possibility of monitoring all these stages in real-time. 
  
\end{abstract}

\section{Introduction}
Surface wettability at the nanoscale still poses many questions: the processes ultimately governing the evaporation of sessile nanodroplets are incompletely known, while the final stages of their evaporation down to an extended molecular film is little documented. This insufficient knowledge persists despite remarkable advances in micro and nano-fluidics techniques, and despite numerous applications in nano-printing, spray cooling and liquid nano-dispensing, which spark interest for these fundamental questions. At the nanoscale not only the production and delivery \cite{Ondarcuhu2011} but also the real-time imaging and analysis of droplets remain outstanding challenges, limiting our capacity of investigation. The evaporation time of nano-droplets is expected to be shorter than their macroscopic counterparts, requiring fast probes to track their evolution, while a very high sensitivity is required to measure their miniature mass. In order to access intrinsic phenomena, the observer should additionally not perturb the droplet evolution while measuring it. There is today no probe technology for nanodroplets meeting all these requirements at once.

Droplets smaller than few femtoliters are difficult to track by optical imaging, which faces the resolution limit imposed by diffraction. Atomic Force Microscopy (AFM) and Transmission Electron Microscopy (TEM) can break this limit and have proved to be powerful tools to analyze nanoscale droplets, however with restrictions. Non-contact AFM has been employed to analyse the shape of sessile droplets of a few microns of diameter and less with minimal perturbation \cite{Checco2006}, however at smaller volume these experiments are increasingly difficult. Recent works have employed a two-dimensional material covering a substrate (for example graphene on mica) in order to extend AFM measurements to yet smaller droplets of few tens of nanometers diameter \cite{Heath2011,Dong2014}, however in this case it becomes difficult to disentangle interactions between the substrate, the liquid and the 2D material. Spectacular results were obtained as well by a non-contact extension of AFM dubbed scanning polarization force microscopy (SPFM), which achieved lateral resolution of 10 nm to study droplets on pristine surfaces \cite{Salmeron95,Rieutord1998a}. Environmental TEM has also allowed direct visualization of sessile nanodroplets with diameters of tens of nanometers, however in these experiments the incident electron beam leads to unavoidable charging of the droplets, whose shape and dynamic evolution become ruled by electrostatic effects \cite{Mirsaidov2012, Leong2014, Yang2019}. Unfortunately, all of these microscopy techniques are inherently slow, with typical time resolution of a minute. The fast evaporation dynamics of nanodroplets has thus remained experimentally out of reach.

Micro and nanomechanical mass sensors \cite{Yang2006, Sage2015} have been applied to the investigation of small liquid volumes as well, by looking at mechanical resonance shifts induced by the liquid adsorbate, in the spirit of quartz microbalances \cite{Rodahl1995} but with improved sensitivity \cite{Arcamone2007,Golovko2009,Park2012,Prasad2014}. In \cite{Arcamone2007} notably, the time evolution of evaporating glycerol droplets of 1 to 5~$\mu$m of diameter (1.5-150~aL) could be tracked with a time resolution of 10 s and a mass sensitivity of 64~fg, showing an evaporation behavior consistent with that prevailing at the macroscopic scale. In order to open investigations at the nanoscale, and to track the much faster dynamics of infinitesimally small liquid volumes, we propose here a new approach. We make use of a miniature optomechanical disk resonator that supports co-localized optical and mechanical modes. As a result of it's small dimensions, ultra-high mechanical frequencies and efficient optomechanical transduction \cite{Favero2009,Aspelmeyer2014}, the resonator offers a sensitivity to punctual mass adsorption of 100~ag (100~zL) while ensuring a time resolution of a millisecond. On top of providing mechanical measurement of the nanoscale liquid adsorbate, the device allows extracting optical information in parallel, leading the possibility of a multiphysics analysis. With this platform, we achieve real-time tracking of single droplets of 2-propanol liquid that evaporate from a micrometric to a nanometric size over a few milliseconds, all the way down to attolitter volumes, where a transition to an extended film is observed. The multiphysics signals enable to unambiguously determine the geometrical evaporation mode. The fast evaporation dynamics is then modeled with minimal assumptions: as a consequence of their strong curvature, sessile nanodroplets evaporate even when the surrounding gas is saturated, a phenomenon expected from the Kelvin effect \cite{Kelvin,KelvinNature79,Porstendorfer1977a} but that had not been observed yet. Our millisecond-resolved measurement of an individual nanodroplet enables a direct observation of this nanoscale phenomenon.

\section{Results and discussion}

\begin{figure*}[h!]
	\includegraphics[width=0.9\textwidth]{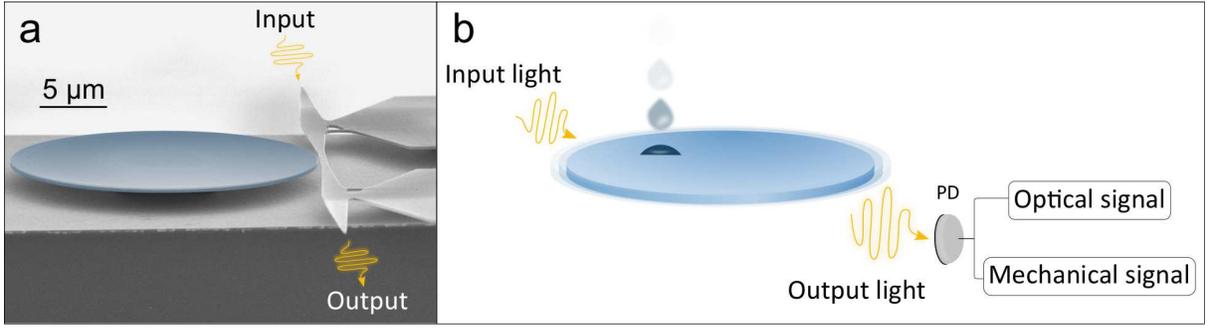}
	\caption{Optomechanical disk for nanodroplet measurement. (a) False-coloured SEM image of the optomechanical resonator employed in these experiments. An integrated waveguide (right-hand side, light grey) is evanescently coupled to the disk (blue). It enables injection (collection) of light into (from) it. (b) Illustration of droplet detection with an optomechanical disk resonator. The DC and RF components of the output optical signal contain respectively information about the interaction of the droplet with the optical and mechanical modes of the disk.}
	\label{fig:Figure1}
\end{figure*}

A scanning electron microscope (SEM) image of our optomechanical device is shown in Figure~\ref{fig:Figure1}a. It consists of a Gallium Arsenide (GaAs) disk of 11~$\mu$m radius and 200~nm thickness sitting on an $1.8$~$\mu$m Aluminum Gallium Arsenide (AlGaAs) pedestal. The optical Whispering Gallery Modes (WGMs) supported by the device can be excited via an integrated waveguide placed in vicinity, which is itself input by a tunable telecom laser. On top of these optical modes, the disk supports co-localized in-plane mechanical modes, providing intense optomechanical coupling between both \cite{Baker2014b}. The input light can be modulated close to a mechanical eigenfrequency, while the output light is demodulated, providing efficient actuation of the disk mechanical displacement, and detection of its amplitude and phase \cite{Sbarra2021}. The co-existence of optical and mechanical modes in the structure provides a dual detection approach for an analyte that would land onto the disk. In our experiments, this landing analyte is additionnally imaged in real-time thanks to a $\times$100 magnification microscope objective coupled to a fast camera with 1~kHz frame acquisition rate. The detection protocol of sub-micron 2-propanol droplets is illustrated in Figure~\ref{fig:Figure1}b. After being generated from a liquid solution by an ultrasonic piezo-ceramic nebulizer, individual droplets softly land onto the resonator surface, perturbing both its optical and mechanical modes. Signatures of these interactions can be retrieved by analyzing the DC and RF components of the output optical signal, respectively. In the followings, we make use of both these optical and mechanical signatures in a dual and complementary manner.\\

\begin{figure}[h!]
	\includegraphics[width=8.5cm]{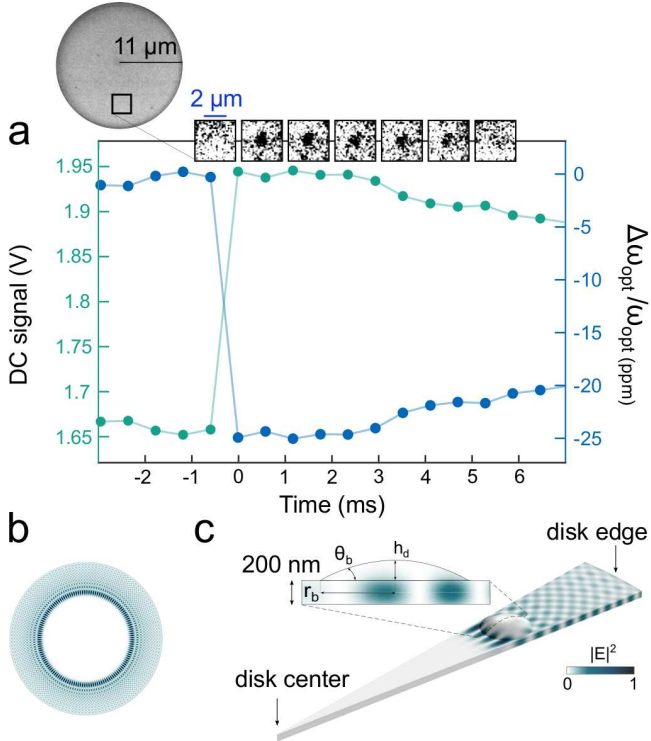}
	\caption{Optical measurement of the landing droplet trough the disk WGM. (a) Top: Frames acquired with our fast camera, extracted from the top-view of the disk (left). Bottom: Experimental optical data acquired during the landing of a single droplet on the resonator. The voltage of the photodiode changes abruptly at t=0 as consequence of the landing. (b) Electric field intensity for a (TE) WGM at 1.55~$\mu$m wavelength with azimuthal order m=68 and radial order p=10, calculated for a GaAs disk of 11~$\mu$m radius and 200~nm thickness. (c) A slice of the disk close to the droplet landing position is shown. Within the droplet volume, the electric field decreases exponentially in the vertical direction. $h_{d}$, $r_{b}$ and $\theta_{b}$ are the height, contact radius and contact angle of the droplet.}.
	\label{fig:Figure2}
\end{figure}

The evanescent part of the WGM in the air surrounding the disk is perturbed by the landing of a droplet, and the optical resonance frequency red-shifted in consequence.The larger the WGM field amplitude within the droplet, the larger the resonance shift. This optical sensing approach has been extensively used to measure solid adsorbate particles \cite{Vollmer2008a, Zhu2010, He2011, Zhang2015}, while few examples exist for adsorbed liquids \cite{Qiulin2010}. We use it here to measure our landing droplets. Figure~\ref{fig:Figure2}a reports the DC photodiode voltage (left axis) acquired upon landing of a single 2-propanol droplet on our optomechanical resonator. This quantity can be linearly converted into an optical resonance shift (right axis), thanks to the knowledge of the thermo-optic optical response of our device, in a blue-detuned regime where we are operating here \cite{Parrain2015}. The frames captured by the fast camera are shown in correspondance of the time-axis (top of Figure~\ref{fig:Figure2}a), allowing real-time top-view visualization of the landing position of the sessile droplet, while enabling a rough estimation of the droplet base radius. At time t=0, the relative resonance shift is about -20 ppm and the base radius is estimated to be of 650~nm. With these two information, and assuming a spherical geometry for the droplet, the WGM optical field can be computed in presence of the droplet via the Finite Element Method (FEM). For the (TE) WGM of radial order p=10 and azimuthal order m=68 employed in this measurement, whose top-view distribution is shown in Fig.~\ref{fig:Figure2}b, we obtain the vertical field distribution reported in Fig.~\ref{fig:Figure2}c. Using an analytic formula obtained from a perturbation theory \cite{Teraoka2006} to retrieve the resonance shift 
\begin{equation}
\frac{\Delta \omega_{opt}}{\omega_{opt}}=-\frac{1}{2} \frac{  \int_{V_d} \delta\epsilon_{r} E_{0}^{*}(r) E_{d}(r) dr}{\int_{V_{WGM}} \epsilon_{r} E_{0}^{*}(r) E_{0}(r)  dr} 
\end{equation}
with $V_d$ the droplet volume, $\epsilon_{r}$ the relative dielectric permittivity, $E_{0}$ the unperturbed WGM field, and $E_{d}$ the field in the droplet, taken equal to $E_{0}$ at first order, we deduce a droplet contact angle $\theta_{b}$ of 12$^\circ$ at t=0, corresponding to a liquid volume of 46~aL. The main source of uncertainty stems from the limited resolution of our imaging system (about 100 nm), which affects our precision on both droplet radius and radial and azimuthal landing position, eventually leading to a 20 $\%$ uncertainty in the estimation of the droplet volume.\\

The evaporation of the droplet is investigated by analyzing the video frames and optical signal at t>0, in order to distinguish evaporation in the constant contact radius (CCR) and constant contact angle (CCA) modes \cite{Arcamone2007,Golovko2009,Park2012,Prasad2014}. In our experiment, the base radius remains unchanged during the first three frames after landing ($<3~$ms), while the photo-diode signal remains flat, showing no evident evaporation. Subsequently, the relative optical shift starts decreasing in amplitude, without reaching back to its initial value. At the same time, there is no droplet visible anymore in the video frames ($\sim6~$ms), which is consistent with a strong reduction of the droplet height. The relative optical shift evolution between 0 and $7~ms$ can be reproduced by FEM simulations of a spherical droplet using the CCR evaporation mode, while the CCA mode is in contrast excluded by our data. In CCR mode at time t=5~ms, the contact angle is close to 9$^\circ$, which corresponds to a height of 53~nm. In order to exclude hybrid evaporation modes that our imaging system could hardly distinguish, we need to confront this CCR interpretation to an independent set of data. To that purpose, we make use of a powerful feature of our experimental approach: the possibility of dual optical and mechanical sensing by the resonator.

\begin{figure}[H]
	\includegraphics[width=8.5cm]{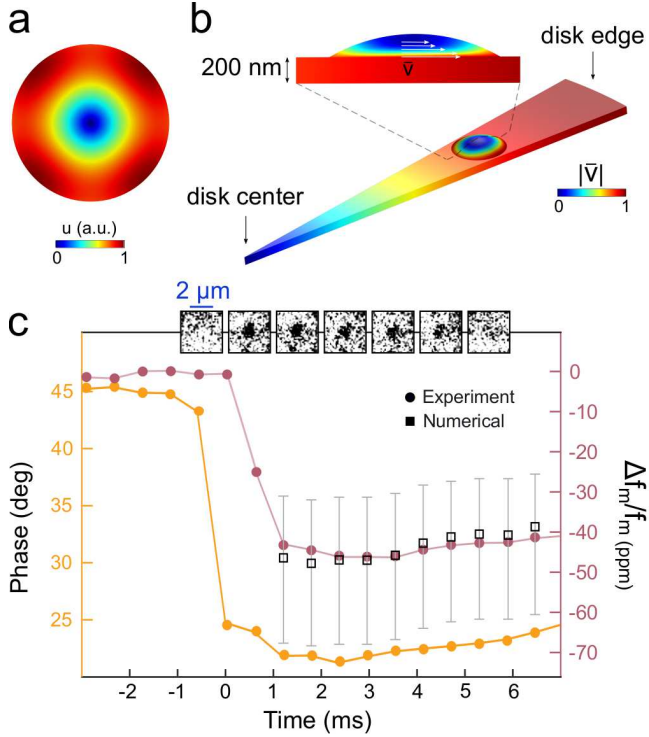}
	\caption{Mechanical measurement of the nanodroplet by the disk. (a) Displacement profile of the RBM1 for a disk of 11~$\mu$m radius and 200~nm thickness. (b) RMS velocity amplitude distribution for the same RBM1 over a disk slice where the droplet did land. The in-plane vibrations of the disk induce shear-waves in the droplet that are attenuated along the vertical direction, while the no-slip condition is imposed at the solid-liquid interface. For the sake of illustrating the situation at t=0, the droplet contact angle was artificially increased from 12$^\circ$ to 35$^\circ$. (c) Phase of the experimentally measured output signal demodulated at the RBM1 frequency (yellow), acquired during the droplet landing already reported in Fig.\ref{fig:Figure2}a. The phase changes abruptly at time t=0 when the droplet lands. The mechanical resonance shift (after correction, see text) is reported in pink. A 1~kHz bandwidth is employed as a compromise between sensitivity to mass adsorption (hundred of attograms) and time-resolution (0.6~ms). The square open symbols are the mechanical resonance shifts calculated with FEM (see text). The associated error bars account for the uncertainty in the droplet contact radius ($\pm{200}$ nm) and position on the disk surface ($\pm{200}$ nm).}
	\label{fig:Figure3}
\end{figure}

We now look at the RF signal associated to the mechanical frequency shift of the fundamental radial breathing mode (RBM1) of the resonator, which resonates at $f_m=130$~MHz. The displacement profile of RBM1 is reported in Figure~\ref{fig:Figure3}a: it consists of an in-plane radial vibration, which can generate shear waves inside a liquid droplet deposited on the disk top surface. Because of the liquid viscous response, these waves are attenuated over the penetration depth $\delta=\sqrt{\frac{\eta}{\pi \rho f_m}}$, with $\eta$ the dynamic viscosity and $\rho$ the density, as illustrated in Figure~\ref{fig:Figure3}b. Just like a shear wave sensor \cite{Prasad2014}, our disk mechanical resonator hence senses a limited amount of the deposited liquid (at 130~MHz, $\delta\sim$50~nm in water and $\delta\sim$80~nm in 2-propanol). When the liquid thickness is comparable to $\delta$, shear waves are reflected at the top liquid-air interface and must be accounted for\cite{Ondarcuhu2013}. For these reasons, calculating the RBM1 mechanical resonance shift induced by a landed droplet is achieved with a numerical approach that solves the disk elasticity problem together with the droplet Navier-Stokes equations, imposing a continuity of the tangential velocity at the solid-liquid interface (no-slip condition). In our FEM calculations we neglect the effect of surface tension, which only becomes relevant for radius of curvature of $\sim$10~nm. Figure~\ref{fig:Figure3}b shows calculations carried in the slice of the disk where the droplet analyzed in Figure~\ref{fig:Figure2}a did land, revealing a RMS velocity amplitude that decreases towards the top of the droplet, while its bottom is shaken by the disk motion. The mechanical frequency shift is maximized if the droplet lands at the disk periphery, where the disk velocity is larger, or if it spreads on the surface with a smaller contact angle and larger contact radius.

The mechanical shift measured in response to the droplet landing is reported in Figure~\ref{fig:Figure3}c (right axis), over the same time interval as Figure~\ref{fig:Figure2}a, together with the phase of the output signal demodulated at the RBM1 frequency (left axis). The latter is converted into the former in the linear region of the phase-frequency response where we drive the resonator, however with a correction that accounts for the direct modification of the mechanical frequency induced by the change of cavity photon number following a landing event. This latter effect is not captured by our FEM mechanical simulations, but it is subtracted from our experimental signal using a prior calibration and a multiphysics model (optical, mechanical, thermal) that describes the actuation/detection of optomechanical resonators \cite{Sbarra2021}. The corrected mechanical shifts reported in Figure~\ref{fig:Figure3}c can be confronted to those predicted by FEM calculations, provided that the geometry of the droplet is known at each time. Using the CCR evaporation geometry deduced from the optical signals reported in Figure~\ref{fig:Figure2}a, we obtain the mechanical shifts shown as open square symbols in Figure~\ref{fig:Figure3}c. The close agreement with experimental data reinforces the interpretation of a CCR evaporation mode for the droplet over the first $7$ ms, before it disappears from our video images.\\

\begin{figure}[H]
	\includegraphics[width=0.43\textwidth]{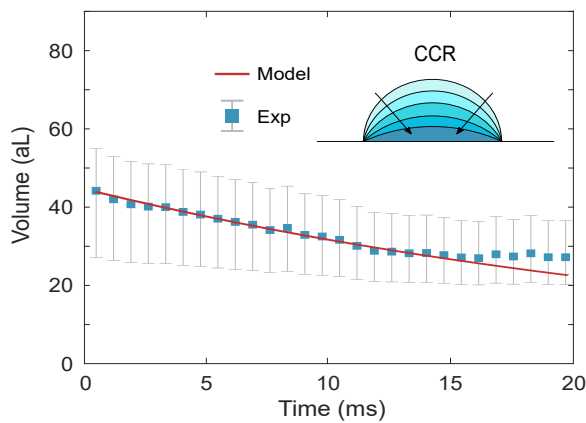}
	\caption{Time evolution of the nanodroplet volume. The data of Figures \ref{fig:Figure2} and \ref{fig:Figure3} are shown in terms of droplet volume, over the first 20 ms, for the elucidated CCR mode. Blue symbols: measurements. Red line: evaporation model including the Kelvin effect (see text). The error bars affecting the estimated volume account for the uncertainty in the droplet contact radius ($\pm{200}$ nm) and position on the disk surface ($\pm{200}$ nm) deduced from the video frames.}
	\label{fig:Figure4}
\end{figure}

Now that the CCR mode of evaporation is established, we turn our attention to the physical mechanism governing the evaporation dynamics. It is a fast evaporation, with a droplet volume of 45 attolitters that reduces by half in 15 ms, as shown in Figure~\ref{fig:Figure4}. As originally expressed in the Langmuir model \cite{Langmuir1918} for a spherical droplet, evaporation is a diffusion process driven by a pressure gradient in the surrounding gas. Hu and Larson \cite{Hu2002} extended this diffusive model to a sessile spherical droplet, and expressed its mass reduction rate:
\begin{equation}
	-\frac{dm}{dt}= \frac{\pi r_{b} D M}{RT} (p_{sat}-p_{\infty})(0.27\times\theta_{b}^{2}+ 1.30)
	\label{eq:HuLarson}
\end{equation} 
with $r_{b}$ the contact radius, $D$ the diffusion coefficient, $M$ the molar mass of the liquid, $T$ the temperature, $p_{sat}$ the saturation pressure and $p_{\infty}$ the partial pressure in the surrounding gas. In the CCR mode, $r_{b}$ remains constant while $\theta_{b}$ and the sphere radius $R_{s}$ evolve. The geometric relations for a spherical droplet $r_{b}=R_{s}\sin\theta_{b}$ and m=$\rho\frac{\pi R_{s}^{3}}{3}(1-\cos\theta_{b})^{2}(2+\cos\theta_{b})$ hold all along the evolution, and enable integrating Eq.\ref{eq:HuLarson} on a numerical solver. At room temperature, for the parameters of 2-propanol and for the initial $r_{b}=650$ nm observed in our experiments, the calculation predicts a reduction in volume by a factor two in $70$ $\mu$s if $p_{\infty}=0.9 \times p_{sat}$ and in $0.7$ ms if $p_{\infty}=0.99\times p_{sat}$. These timescales are much shorter than that observed in our experiments, hence we deduce that $p_{\infty}\approx p_{sat}$, which confirms that the mist produced by the nebulizer saturates the environment of the resonator. Under this condition, Eq.\ref{eq:HuLarson} does not predict evaporation to take place. The missing ingredient is the Kelvin effect linked to the strong curvature of the droplet \cite{Kelvin}, which replaces the saturation pressure in Eq.\ref{eq:HuLarson} by $p_{sat}^{K}=p_{sat}\times \exp{ \frac{2 \gamma M}{RT\rho R_{s}} }$, with $\gamma$ the liquid surface tension. With this ingredient included, we are able to reproduce the measured dynamics of evaporation with no adjustable parameter, as represented by the solid line of Figure~\ref{fig:Figure4}. This constitutes a direct observation of the role of surface tension in the evaporation of a nanodroplet.

Our above analysis is based on the assumption that a sessile droplet in the capillary regime adopts a spherical shape. At the nanoscopic scale, AFM studies revealed that the topography of sub-micron droplets is in contrast distorted by surface inhomogeneities and the presence of long-range forces \cite{Tamayo1996,Rieutord1998,Checco2006,Ondarcuhu2013,Starov2019}. In such cases, a transition zone close to the contact line ensures the continuity of hydrostatic pressure between the spherical part of the droplet and an underlying thin film of liquid constituted of molecules continuously adsorbing and desorbing on the substrate \cite{Strawhecker2005,Asay2006}. Dynamic molecular simulations do predict the existence of such a regime \cite{Nieminen1992, LiuZhang2013, Pillai2018}, Our present experiments are carried within an environment saturated by the sprayed mist of 2-propanol molecules, hence the sessile droplets we investigate should be accompanied by such extended film \cite{Yakovlev,Gee,Barnette} lying on the resonator surface \cite{Parrain2015}. This thin (precursor) film is essential to understand the wetting of the deposited droplet and the final stages of its evaporation. However, detecting such film with sufficient time-resolution to reveal its dynamics has remained out of reach, owing to the very small volume of liquid involved. Our optomechanical measurement is efficient enough to address this problem, enabling real-time analysis of the last steps of evaporation of a single nanodroplet with millisecond resolution.

\begin{figure*}[h!]
	\includegraphics[width=0.9\textwidth]{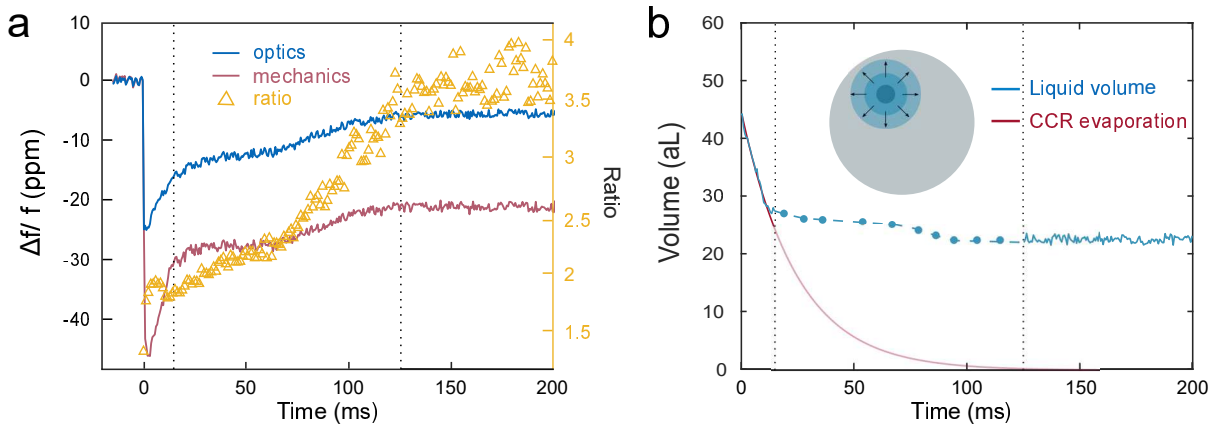}
	\caption{Optomechanical measurement of the transition from a nanodroplet to an extended film. (a) Optical (blue) and mechanical (red) normalized resonance shift during the 200~ms after droplet landing. The two signals do not reach back to their initial value at the end, indicating that a residual amount of liquid is still present on the disk surface. Their ratio is reported with yellow triangles: it remains constant at 2.5 during the first 5~ms, consistent with a localized droplet of spherical shape, and then increases up to 3.5, consistent with the stabilization of an extended liquid film on the disk surface. (b) The detected volume as function of time of  liquid is finally deduced by FEM simulations (see main text). It consists first of the CCR evaporation stage governed by the Kelvin effect, followed by the spreading of a puddle precursor film illustrated in blue (at t= 54 ms, puddle diameter= 3.46 $\mu$m, height=2.7 nm; at at t=104 ms, puddle diameter= 6.58 $\mu$m, height=0.66 nm), and ends with the stabilization of an extended film on the surface.}
	\label{fig:Figure5}
\end{figure*}
To investigate the evolution of the spherical droplet and its co-existence with an extended film, we take advantage of our ability to perform dual optical and mechanical sensing experiments in real-time. We analyse concomitantly the normalized optical and mechanical resonance shifts for the same landing event as above, but over a longer time interval of a fraction of second, and track the ratio of the two shifts (Figure~\ref{fig:Figure5}a). This ratio displays some regularity, evolving from a plateau around a value of 2.5 (during the first 7~ms) to a plateau around 3.5 (130 to 200 ~ms after droplet landing), while optical and mechanical signals do not recover their original value over this period. The value of 2.5 can be reproduced by FEM considering a spherical droplet localized at the identified landing position, consistent again with our above interpretation of the first stages of evaporation. At longer time, the conclusions are very different: the value of 3.5 is not consistent with a spherical droplet, whatever its dimensions, but it is in contrast reproduced by considering an extended film of 2-propanol enveloping the disk resonator. For a thin film of thickness $t<<\delta$, the attenuation of the shear velocity and electric field intensity in the direction perpendicular to the disk surface can be neglected, and the mechanical frequency shift takes the simple form $\Delta f_m/f_m=-\frac{t}{2}\frac{\rho \int_{S} u^2(r) dS}{m_{eff}}$ with $\rho$ the density of the liquid, $m_{eff}$ the resonator effective mass, $u(\mathbf{r})$ the normalized mechanical displacement amplitude and $\int_{S} dS$ the integral over the disk surface. In this case, the ratio $R$ of normalized mechanical and optical shifts becomes independent of t:
\begin{eqnarray}
	R=\frac{\rho \int_{S} u^2(r) dS}{m_{eff}} \frac{\int_{WGM} \epsilon_r\vert E_{0}(r)\vert ^2 dV }{\int_{S} \delta\epsilon_{r} \vert E_{0}(r)\vert ^2 dS}
	\label{eq:Experim_dropl_longTime}
\end{eqnarray}
and adopts the numerical value 3.5 for the specific optical and mechanical disk modes employed in our sensing experiment. The robust experimental observation that this ratio converges towards 3.5, for multiple independent droplet landing events, is a strong indication that an extended thin film is indeed formed at the surface of the resonator in the final stages of the evaporation. From the value of optical and mechanical shifts at t>130 ~ms, we estimate an accumulated amount of liquid of 22~aL (Figure~\ref{fig:Figure5}b). If uniformly distributed over the disk surface, this would correspond to a film thickness of less than 1~\AA, consistent with the thin film condition. Between the end of the initial evaporation ($\sim$7~ms) and the final thin film regime (>130~ms), we anticipate a hybrid regime where the evaporating droplet evolves into a puddle of growing diameter. In Figure~\ref{fig:Figure5}b, we model this regime by approximating the droplet as a cylindrical puddle of varying diameter and height. At each time, there is a unique diameter that enables retrieving the proper value of R, and a unique height that allows retrieving the observed individual shifts. This minimal model enables estimating the detected liquid volume at each time (blue symbols). While simple, it shows that the residual liquid volume of $\sim$25~aL is spreading in time over the surface with a somewhat irregular dynamics, probably a consequence of the presence of pinning centers on the surface.

\section{Conclusions}
Miniature optomechanical resonators appear here as a new tool to study wetting and evaporation dynamics of nanoscale liquid volumes with high temporal resolution. The high sensitivity of our dual optical/mechanical sensing device allows detection with millisecond resolution of a liquid volume in the range of tens of attoliters, either localized (within a spherical droplet) or distributed over the disk surface (in the form of a thin film). This enables us tracking the fast dynamics of a nanodroplet  evaporating in a saturated environment, a situation that had been theoretically discussed and believed to play a role in the evolution of aerosols \cite{Porstendorfer1977a}, but which had not been measured at the single droplet level because of a lack of suitable techniques. Beyond this first original outcome, the high sensitivity and time-resolution of optomechanical techniques open to several interesting problems in nanoscale fluidics, such as fast thermodynamic transitions to solid and glass states in nanoscale droplets, or the response of a nanoscale droplet to rapidly-evolving boundary conditions at liquid-solid interfaces.

\section*{Methods}
\subsection*{Experimental setup}
The telecom wavelength laser is coupled to the waveguide with two micro-lensed fibers for injection and collection. The amplitude of the laser is modulated at a frequency close to the mechanical resonance (130 MHz, $Q_{m}\sim 10^4$) with a Mach-Zehnder electro-optic modulator. The modulation of light induce photothermal forces in the disk that actuates its mechanical motion. A lock-in amplifier (UHFLI 600 MHz, Zurich Instruments) is used to demodulate the output signal and read its amplitude and phase in real time. Using the phase-frequency characteristic of the resonator, its mechanical frequency is directly retrieved. More details on the multiphysics processes that come into play in the actuation and detection schemes can be found in \cite{Sbarra2021}.
 
\subsection*{Numerical modelling}
The optical and mechanical modes observed experimentally were identified using COMSOL Multiphysics, a 3D finite element model software. We also used a multiphysics model to establish the impact of a sessile droplet on the optical and mechanical resonance frequencies of the resonator with a no-slip boundary condition at the interface between the liquid sessile droplet and the disk. The results are used to retrieve the geometry parameters (contact angle, base radius, volume) of the droplet from the experimental data.

%\begin{suppinfo}
%The following files are available free of charge.
%\begin{itemize}
%  \item Filename: brief description
%  \item Filename: brief description
%\end{itemize}

%\end{suppinfo}

\bibliography{library}

\section{Acknowledgement}
The authors acknowledge support from the European Commission through the VIRUSCAN (731868) FET-open and NOMLI (770933) ERC projects, and from the Agence Nationale de la Recherche through the Olympia and QuaSeRT projects.
\section{Author contributions}
S. Sb, L. W and I. F planned experiments, analyzed data, developed models and wrote the article. S. Sb and L. W. carried experiments and fabrication of resonators. S. Su optimized the clean-room processes and provided assistance to fabrication. A. L. grew the epitaxial wafers employed for resonators. 
\section{Competing interests}
The authors declare to have no competing interest.

\end{document}